\UseRawInputEncoding
\documentclass[11pt,a4paper]{article}
\usepackage{jcappub}

\usepackage{aasmacros}

\usepackage[english]{babel}               

\usepackage{lmodern}        
\usepackage[T1]{fontenc}    
\usepackage{mathpazo}                                            

\usepackage{tabu}

\usepackage{geometry}     
\usepackage[final, babel]{microtype} 

\usepackage[textwidth=2.4cm]{todonotes}%
\usepackage{doi}
\usepackage{hyperref}
\hypersetup{
  colorlinks=true,        
  linkcolor=blue,         
  citecolor=cyan,         
}

\usepackage{graphicx}
\usepackage{dcolumn}
\usepackage{bm}
\usepackage{color}

\usepackage{amsmath}
\usepackage{amssymb}
\usepackage{float}

\usepackage{soul,xcolor}
\setstcolor{red}

\title{Extending the weak cosmic censorship conjecture to the charged Buchdahl star by employing the gedanken experiments}

\author[a,b,c,d]{Sanjar Shaymatov}
\author[e]{and Naresh Dadhich}

\affiliation[a]{Institute for Theoretical Physics and Cosmology, Zheijiang University of Technology, Hangzhou 310023, China}
\affiliation[b]{Akfa University,  Milliy Bog Street 264, Tashkent 111221, Uzbekistan}
\affiliation[c]{Institute of Fundamental and Applied Research, National Research University TIIAME, Kori Niyoziy 39, Tashkent 100000, Uzbekistan}
\affiliation[d]{National University of Uzbekistan, Tashkent 100174, Uzbekistan}
\affiliation[e]{Inter University Centre for Astronomy \&
Astrophysics, Post Bag 4, Pune 411007, India}

\emailAdd{sanjar@astrin.uz, nkd@iucaa.in}
\date{\today}
\abstract{In this paper, we wish to investigate the weak cosmic censorship conjecture (WCCC) for the non black hole object, Buchdahl star and test its validity. It turns out that the extremal limit for the star is over-extremal for black hole, $Q^2/M^2 \leq 9/8 >1$; i.e., it could have $9/8 \geq Q^2/M^2 > 1$. By carrying out both linear and non-linear perturbations, we establish the same result for the Buchdahl star as well. That is, as for black hole it could be overcharged at the linear perturbation while the result is overturned when the non-linear perturbations are included. Thus WCCC is always obeyed by the Buchdahl star. }


\begin{document}

\maketitle


\section{Introduction}
\label{introduction}

In general relativity, Penrose first proposed  the singularity theorem in 1965~\cite{Penrose65a} which was further generalized into the powerful Penrose-Hawking singularity theorems~\citep{Hawking-Penrose70}. The emergence of singularity marks the limit of validity of a theory, and so is the case for Einstein's theory of gravitation, General relativity (GR). Although it is the most successful theory, yet it is  incomplete as indicated by occurrence of singularities~\citep{Hawking-Penrose70}. Later on, in 1969 Penrose proposed~\citep{Penrose69} the weak cosmic censorship conjecture (WCCC) stating that gravitational collapse would always result in forming an event horizon of black hole which would hide the singularity inside. That is, there is a cosmic censor prohibiting occurrence of naked singularity in gravitational collapse. A singularity would therefore be  always hidden behind the black hole horizon and would never be visible to outside observer. However it is a conjecture and there is no theorem proving its validity. There has been intense activity of testing this conjecture by constructing counter examples~\cite[see, e.g.][]{Christodoulou86,Joshi93,Joshi00,Goswami06,Harada02,Stuchlik12a,Vieira14,Stuchlik14,Giacomazzo-Rezzolla11,Joshi15}]. Despite this, the jury is still out whether naked singularities do or do not form in gravitational collapse. That is, the question of validity of WCCC remains  open.

Another aspect of the question is, could an existing black hole horizon of charged and/or rotating black hole be destroyed by over-extremalizing (i.e., $Q^2>M^2$ for a charged black hole)  . That is, could a black hole be over-extremalized so that singularity is laid bare -- naked. The question was
first addressed by Wald~\cite{Wald74b}. He showed that an extremal black hole could never be extremalized by a test particle accretion. Further it was also shown that a non-extremal black hole cannot be converted into an extremal one~\cite{Dadhich97}. What happens is that as extremality approaches, the parameter window for accreting  particle hitting the horizon pinches off. That is once a horizon is formed, it cannot be destroyed and WCCC cannot be violated.

Later, Hubeny \cite{Hubeny99} showed that a  black hole can indeed be overcharged in a discontinuous process involving linear order perturbations. That is, extremality cannot be attained but it could perhaps be jumped over in a discrete process. The above experiment was also applied to test the WCCC for Kerr and Kerr-Newman black holes \cite{Jacobson09,Saa11}. The horizon of black hole can be destroyed provided in-falling  particle adds enough electric charge/angular momentum to black hole's charge/angular momentum. After that, a large number of works have been done in various contexts~\cite[see, e.g.][]{Shaymatov15,Bouhmadi-Lopez10,Rocha14,Jana18,Song18,Duztas18,Duztas-Jamil18b,Duztas-Jamil20,Yang20a,Yang20b} initially  without including  backreaction effects. Later, it was also shown that inclusion of back reaction effects did not alter the earlier result~\cite[see,e.g.][]{Zimmerman13,Rocha11,Isoyama11,Colleoni15a,Li13,Shaymatov19b}. Also it is worth noting that the above experiment was extended to the various black hole solutions, i.e.asymptotically AdS black holes~\cite{Gwak16,Natario16,Natario20,Zhang14,Gwak16PLB}.

Although this experiment gave rise to intense discussion on violation of the WCCC, Sorce and Wald \cite{Sorce-Wald17} once again put it all to rest by showing that when second order perturbations are taken into account, the WCCC that was violated at the linear order would always be restored.  So a new version of the gedanken experiment has been proposed, with inclusion  non-linear  perturbations ~\cite{An18,Gwak18a,Ge18,Ning19,Yan-Li19,Shaymatov19a,Shaymatov19b,Jiang20plb}, the linear order result of WCCC violation is always overturned.  The new version of the gedanken experiment has been applied to black holes in higher dimensions.
It was already shown that higher dimensional charged black hole could be overcharged for linear order perturbations ~\cite{Revelar-Vega17}. However, five dimensional charged rotating black hole having single rotation cannot be over-extremilized when rotation parameter dominates over charge~\cite{Shaymatov19c}. However it turns out that in  $D\geq6$~\cite{Shaymatov22JCAP,Shaymatov21a,Shaymatov19a,Shaymatov20a} it cannot be overcharged even at the linear order accretion. Note that the  gedanken experiment has also been recently extended to RN-AdS black hole \cite{Wang20} and RN-dS black hole surrounded by dark matter field~\cite{Shaymatov21d} under the non-linear order perturbation.

In this paper we wish to extend this WCCC analysis to the charged Buchdahl star. It is defined \cite{Dadhich22,Pani22} by gravitational potential, $\Phi(r) = 4/9$ while black hole by $\Phi(r)=1/2$. The Buchdahl compactness limit is given by $\Phi(r) \leq 4/9$. The Buchdahl star is the most compact non black hole object having the limiting compactness bound, {
and it is the most compact stable object~\cite{Hod18,Dadhich20:JCAP}. Thus, it should be pertinent and interesting to extend the weak censorship conjecture to Buchdahl star by applying a new version of the gedanken experiments.}  Note that the characterization, $\Phi(r) = 1/2, 4/9$ respectively for black hole (BH) and Buchdahl star (BS) is universal irrespective of  object being  uncharged and non-rotating or otherwise. It is remarkable  that BS shares almost all BH properties \cite{Sum-Dad22} including, in particular, the  extremal limit which for the charged case is $Q^2/M^2>9/8>1$. So an extremal BS is over-extremal relative to BH. If quantum mechanical corrections are included, it might be possible to have a solution with the horizon, allowing charge $Q$ that is slightly greater than $M$ ($Q < \sqrt{2} M$) as shown in \cite{Casadio15:PLB}. However, inclusion of quantum corrections are beyond the scope of the present study. We shall study both linear and non-linear perturbations of BS in the context of WCCC and show that the result is the same as for BH. That is, it is possible to violate WCCC at the linear order which is then overturned when non-linear perturbations are included.

The one critical difference for the Buchdhal star is that its boundary is timelike as against the null horizon for black hole. For the former, perturbations could be reflected back or flow out as timelike boundary could be crossed both ways. This would further restrict the overcharging process and hence would not however conflict with the result that the Buchdahl star cannot be overcharged.

The paper is organized as follows: In section~\ref{Sec:BS} we briefly discuss the Buchdahl star metric and its properties. In section~\ref{Sec:IW} we describe variational identities and Einstein-Maxwell theory for linear and non-linear variational inequalities. In section~\ref{Sec:Gedanken} we build up linear and non-linear order perturbations inequalities for studying overcharging of Buchdahl star. Finally, we end with a discussion in section~\ref{Sec:Discussion}.

\section{The Buchdahl star space-time metric}\label{Sec:BS}

The spherically symmetric Reissner-Nordstr\"{o}m metric describes gravitational field of a charged static object whether black hole or otherwise.  It is given by
\begin{eqnarray} \label{Eq:metric}
d s^2 &=& \, -F(r) dt^2 + F(r)^{-1}\, dr^2  +  r^2d\Omega^2\, ,
  \end{eqnarray}
with the line element of 2-sphere $d\Omega^2$ and
\begin{eqnarray}
 F(r)&=& 1- 2\Phi(r)\, ,
 \end{eqnarray}
where $\Phi(r) = (M-Q^2/2r)/r$ with $M$ and $Q$ corresponding to  mass and electric charge of BS. {
A black hole is defined by the condition $\Phi(r=r_{+}) = 1/2$ that gives the black hole horizon, while Buchdahl star by $\Phi(r=r_{BS})=4/9$ that gives the surface, radius of the star \cite{Dadhich22,Pani22}.} The Buchdahl compactness bound is in general given by $\Phi(r)\leq4/9$, \cite{Dadhich20:JCAP}. BS is defined by the equality indicating the most compact non black hole object.

{The BS condition,
 \begin{eqnarray}
\Phi(r=r_{BS}) = (M-Q^2/2r)/r = 4/9\, ,
\end{eqnarray}
gives
\begin{equation}
\frac{M}{r_{BS}}
=\frac{8/9}{1+ \sqrt{1-(8/9){\beta^2}}} \, ,
\end{equation}
or BS surface radius is given by
\begin{equation}
r_{BS} =\frac{9M}{8}(1+\sqrt{1-(8/9){\beta^2}}) \, ,
\end{equation}
where $\beta^2=Q^2/M^2$.} 
{Here and in what follows, $r_{BS}$ would indicate BS surface radius.} We would use the same symbol, {$r_{+}=r_{BS}$}, as for BH, but it would be clear from the context, in particular the appearance of the factor $8/9$ would clearly identify BS.
Thus the extremal limit for BS is defined by $\beta^2=9/8 >1$, which is over extremal for BH ($r_{+}=M(1+\sqrt{1-\beta^2})$). It is interesting to note that BS could be overcharged relative to BH.

The electromagnetic potential is given by
\begin{eqnarray} \label{el:pot} \textbf{A} = -\frac{Q}{r} dt \, . \end{eqnarray}
Because of spherical symmetry, it is straightforward to compute area, surface gravity and electric potential  at any $r$, and so at the BS boundary as well. In particular, the surface gravity is given by
\begin{equation}\label{Eq:k}
k = \frac{F'(r)}{2}|_{r=r_{BS}} \, = \frac{M}{r_{BS}^2}\sqrt{1-(8/9)\beta^2}\, .
\end{equation}
Note that for BH surface gravity, write $(8/9)\beta^2 \to \beta^2$ and $r_{BS} \to r_{+}$ in the above expression.
 In the next section we study variational identities and perturbation inequalities for linear and non-linear order perturbations for testing  WCCC for the Buchdahl star.

\section{Variational identities and Einstein-Maxwell theory }\label{Sec:IW}

We now adapt the Iyer-Wald formalism to derive perturbation inequalities for linear and non-linear order perturbations under the spherically symmetric perturbation. This formalism bases on a diffeomorphism covariant theory for manifold $\mathcal M$  of any given  dimension. This theory can be identified by the Lagrangian $\mathbf{L}$ including $g_{\alpha\beta}$ and $\psi$ that respectively describe the gravitational field and other fields that exist in the environment surrounding the object~\cite{Wald94,Sorce-Wald17}. So,  it is then possible to consider $\phi=(g_{ab},\psi)$ as dynamical fields around compact objects. {A variation of the Lagrangian $\mathbf{L}$} including all these properties can be written as follows:
\begin{eqnarray}\label{Eq:Lagrangian}
\delta \mathbf{L}=\mathbf{E} \delta\phi+d\bm{\Theta}(\phi,\delta\phi)\, .
\end{eqnarray}
Here $\mathbf{E}$ describes the equations of motion (EOM) that satisfies $\mathbf {E}=0$, whereas $\bm{\Theta}$ is referred to as the symplectic potential. Let us then write the symplectic current including the symplectic potential $(n-1)$-form as

\begin{eqnarray}
{\omega}(\phi,\delta_1\phi,\delta_2\phi)=\delta_1 \bm{\Theta}(\phi,\delta_2\phi)-\delta_2\bm{\Theta}(\phi,\delta_1\phi)\, .
\end{eqnarray}

Next we are able to define the Noether current $(n-1)$-form taking into account the dynamical field, $\phi$, with an arbitrary vector, $\xi^{a}$. So we have
\begin{eqnarray}\label{Eq:NR-charge}
\mathbf{J}_\xi=\bm{\Theta}(\phi,{L}_\xi\phi)-\xi \cdot\mathbf{L}\, .
\end{eqnarray}
%
For $d\mathbf{J}_\xi=0$ the Noether current $(n-1)$-form can be defined by~\cite{Wald95}
\begin{eqnarray}\label{Eq:NR-charge1}
\mathbf{J}_\xi=d\mathbf{Q}_\xi+ \mathbf{C}_\xi\, .
\end{eqnarray}
In the above equation $\mathbf{Q}_\xi$ is referred to as the Noether charge, whereas $ \textbf{C}_\xi=\xi^{a}\textbf{C}_{a}$ marks the above mentioned theory's constraint and can be considered to be zero when $d\mathbf{J}_\xi=0$ continues to hold good for EOM.

We then obtain the linear variational identity for given vector, $\xi^{a}$, on the basis of Eqs.~(\ref{Eq:NR-charge}) and (\ref{Eq:NR-charge1}).  For that  at the Cauchy surface, $\Xi$, we can write
\begin{eqnarray}\label{Eq:first-order}
\int_{\partial\Xi}\delta \mathbf{Q}_\xi-\xi\cdot\bm{\Theta}(\phi,\delta\phi)&=&\int_{\Xi}{\omega}(\phi,\delta\phi,\mathcal{L}_\xi\phi)\nonumber\\&-&\int_{\Xi}\xi\cdot\mathbf{E}\delta\phi-\int_{\Xi}\delta \mathbf{C}_\xi \, .
\end{eqnarray}
From the above equation the variation term for $\xi^{\alpha}$ is the first term on the right hand side $\int_{\Xi}{\omega}(\phi,\delta\phi,\mathcal{L}_\xi\phi)$ that no longer exists as the vector $\xi^{a}$ is taken to be Killing and represents a symmetry of the dynamical field, $\phi$. As long as $\xi^{a}$ behaves as  Killing, it then satisfies EOM, i.e. $\mathbf{E}=\mathcal{L}_\xi\phi=0$. Following the linear variational identity, one can then have the non-linear one at the same surface
\begin{eqnarray}\label{Eq:second-order}
\int_{\partial\Xi}\delta^2 \mathbf{Q}_\xi-\xi\cdot\delta\bm{\Theta}(\phi,\delta\phi)&=&\int_{\Xi}{\omega}(\phi,\delta\phi,\mathcal{L}_\xi\delta\phi)\nonumber\\ &-&\int_{\Xi}\xi\cdot\delta\mathbf{E}\delta\phi-\int_{\Xi}\delta^2 \mathbf{C}_\xi\, .\nonumber\\
\end{eqnarray}

For the Killing vector,  $\xi^{\alpha}$,  Eq.~(\ref{Eq:first-order}) then yields
\begin{eqnarray}\label{Eq:eq-motion}
\int_{\partial\Xi}\delta \mathbf{Q}_\zeta-\zeta\cdot\bm{\Theta}(\phi,\delta\phi)&=&-\int_{\Xi}\delta \mathbf{C}_\zeta\, .
\end{eqnarray}
Here we note that in the above equation one can assume that $\zeta^{a}$ is vector field for which the dynamical field $\phi$ represents the exterior solution of a stationary Buchdahl star. If this is the case $\zeta^{a}=\zeta_{(t)}^{a}$ can then be regarded as the timelike Killing vector field that always satisfies $\mathbf{E}=\mathcal{L}_\zeta\phi=0$ for the EOM. In what follows, we shall restrict ourselves to $\zeta^{a}$ vector.  We further represent the Cauchy surface's boundaries that can start from spatial infinity and continues up to the bifurcation surface, $B$, at the other end. {It is worth noting here that the surface will be of bifurcate type for a near extremal Buchdahl star.} So Eq.~(\ref{Eq:eq-motion}) consists of two boundaries, i.e. spatial infinity at one end and the bifurcation surface at the other, and that is given by
\begin{eqnarray}\label{Eq:surfaces}
\int_{\partial\Xi}\delta \mathbf{Q}_\zeta-\zeta\cdot\bm{\Theta}(\phi,\delta\phi)&=&\int_{\infty}\delta \mathbf{Q}_\zeta-\zeta\cdot\bm{\Theta}(\phi,\delta\phi)\nonumber\\&-&\int_{B}\delta \mathbf{Q}_\zeta-\zeta\cdot\bm{\Theta}(\phi,\delta\phi)\, .\nonumber\\
\end{eqnarray}
Collecting the above results Eq.~(\ref{Eq:first-order}) can be obtained for the linear variational identity in the following form
\begin{eqnarray}\label{Eq:first-order1}
\delta M&=&\int_B[\delta \mathbf{Q}_\zeta-\zeta\cdot\bm{\Theta}(\phi,\delta\phi)]-\int_\Xi\delta \mathbf{C}_\zeta\, .
\end{eqnarray}
where $\delta M$ is referred to as the contribution to the boundary integral at spatial infinity, i.e.
\begin{eqnarray}\label{Eq:infinity}
\int_{\infty}\delta \mathbf{Q}_\zeta-\zeta\cdot\bm{\Theta}(\phi,\delta\phi)&=&\delta M\, .
\end{eqnarray}
What is done in the above is also similar to the case for the non-linear variational identity. Thus, we have
\begin{eqnarray}\label{Eq:second-order1}
\delta^2 M&=&\int_B[\delta^2 \mathbf{Q}_\zeta-\zeta\cdot\delta\bm{\Theta}(\phi,\delta\phi)]\nonumber\\&-&\int_\Xi\zeta\cdot\delta\mathbf{E}\delta\phi-\int_\Xi\delta^2 \mathbf{C}_\zeta+\mathcal{E}_\Xi(\phi,\delta\phi)\, .
\end{eqnarray}
In the above expression, the energy $\mathcal{E}_\Xi(\phi,\delta\phi)$ can be identified by $\delta\phi$ at the Cauchy surface $\Xi$. We will further discuss the terms on the right hand side of  Eq.~(\ref{Eq:second-order1}).

For our purpose we further consider Einstein-Maxwell theory for studying linear and non-linear order perturbations for over-extremalizing Buchdahl star. We write the Lagrangian in the form
\begin{eqnarray}\label{Eq:EM-Lag}
\mathbf{L}=\frac{\bm{\epsilon}}{16\pi}\left(R-F^{\alpha \beta}F_{\alpha\beta}\right)\, ,
\end{eqnarray}
where $\bm\epsilon$ and $F_{\alpha\beta}$ respectively refer to the volume element and Faraday tensor of electromagnetic field. For the dynamical fields, $\phi=(g_{ab},A_{a})$, we write
\begin{eqnarray}
\mathbf{E}(\phi)\delta\phi=-{\epsilon}\left(\frac{1}{2} T^{ab}
\delta g_{a b}+j^a\delta \textbf{A}_{a}\right)\, ,
\end{eqnarray}
with the stress-energy tensor defined by
\begin{eqnarray}
T_{ab}=\frac{1}{8\pi}\left(R_{ab}-\frac{1}{2}g_{ab}R\right)-T_{ab}^{EM}\, ,
\end{eqnarray}
and the current
\begin{eqnarray}
j^a=\frac{1}{4\pi} \triangledown_{b} F^{ab}\, .
\end{eqnarray}
In what follows, we shall write the symplectic potential $\bm{\Theta}$ that is formed from two parts, i.e. electromagnetic and gravity parts. So it takes the following form as
\begin{eqnarray}\label{Eq:symplectic}
\bm{\Theta}_{ijk}\left(\phi,\delta\phi\right)&=&\frac{1}{16\pi}\epsilon_{aijk}g^{ab} g^{cd}(\triangledown_{d}\delta g_{bc}-\triangledown_{b}\delta g_{cd})\nonumber\\&-&\frac{1}{4\pi}\epsilon_{aijk}F^{ab}\delta \textbf{A}_{b}\, ,
\end{eqnarray}
with Levi-Civita tensor $\epsilon_{aijk}$.
The Einstein-Maxwell theory allows the symplectic current to have the form as
 \begin{eqnarray}\label{Eq:sym-current}
\omega_{ijk}&=&\frac{1}{4\pi}\left[\delta_{2}(\epsilon_{aijk} F^{ab}) \delta_{1} \textbf{A}_{b}-\delta_{1}(\epsilon_{aijk} F^{ab}) \delta_{2} \textbf{A}_{b}\right]\nonumber\\&+&\frac{1}{16\pi}\epsilon_{aijk} w^{a}\, .
\end{eqnarray}
This clearly shows  that it consists of the electromagnetic and gravity parts.  The part relevant to the gravity, $w^{i}$, can be written in the following form
\begin{eqnarray}
w^{i}&=&P^{ijkhab}\left(\delta_{2} g_{jk} \triangledown_{h}\delta_{1}g_{ab}-\delta_{1} g_{jk} \triangledown_{h}\delta_{2}g_{ab}\right)\, ,
\end{eqnarray}
where $P^{ijkhab}$ is given by
\begin{eqnarray}
P^{ijkhab}&=&g^{ia} g^{b j} g^{k h}-\frac{1}{2}g^{ih} g^{ja} g^{b k} - \frac{1}{2}g^{i j} g^{k h} g^{ab}\nonumber\\ &-& \frac{1}{2}g^{j k} g^{ia} g^{b h} + \frac{1}{2}g^{jk} g^{i h} g^{ab}\, .
 	\end{eqnarray}
Here, we adopt the following conditions
\begin{eqnarray}
\mathcal{L}_{\xi} g_{ab} &=& \triangledown_{a}\xi_{b} +\triangledown_{b}\xi_{a}\, ,\nonumber\\
\triangledown_{a}\textbf{A}_{b}&=&F_{ab}+\triangledown_{b}\textbf{A}_{a}\, ,
\end{eqnarray}
for the Noether current $J_{\xi}$ to have the following form
\begin{eqnarray}\label{Eq:NC}
(J_{\xi})_{ijk}&=&\frac{1}{8\pi}\epsilon_{aijk } \triangledown_{b}(\triangledown^{[b} \xi^{a]}) + \epsilon_{aijk} T_{b}^{a} \xi^{b}\nonumber\\&+&\frac{1}{4\pi}\epsilon_{aijk}\triangledown_{c}(F^{ca} \textbf{A}_{b} \xi^{b}) + \epsilon_{aijk} \textbf{A}_{b} j^{a} \xi^{b}\, .
\end{eqnarray}
Also the Noether charge $Q_{\xi}$ (see Eq.~(\ref{Eq:NR-charge1}))  can be identified by
\begin{eqnarray}
(Q_{\xi})_{ijk}&=&-\frac{1}{16\pi}\epsilon_{ijk ab}\triangledown^{a}\xi^{b}-\frac{1}{8\pi}\epsilon_{ijk ab} F^{ab } \textbf{A}_{c} \xi^{c}\, ,
\end{eqnarray}
and the constraint of the theory
\begin{eqnarray}
(C_{\zeta})_{ijk}&=&\epsilon_{aijk }(T_{\zeta}^{a} + \textbf{A}_{\zeta} j^{a})\, .
\end{eqnarray}

\section{Perturbation inequalities and gedanken experiment }\label{Sec:Gedanken}

In this section, our aim is to show, as for the charged black hole, whether the results for linear and non-linear order perturbations could also be established for charged Buchdahl star. For that we adopt the new gedanken experiment, proposed by Sorce and Wald \cite{Sorce-Wald17} involving  non-linear  perturbations. For this experiment, we shall consider a one-parameter family of field $\phi(\alpha)$ perturbation for the given spacetime. In doing so, we have
\begin{eqnarray}
G_{ab}(\alpha)&=&8\pi\left[T_{ab}^{GR}(\alpha)+T_{ab}^{EM}(\alpha)\right]\, , \\
\triangledown_{b} F^{ab}(\alpha)&=&4\pi j^a(\alpha)\, .
\end{eqnarray}
Note that $\phi(0)$ satisfies $T_{ab}(0)=0$ and $j^{a}(0)=0$, i.e., infalling particles are supposed to cross the BS boundary. As discussed earlier, the following hypersurface $\Xi=\Xi_{1}\cup H$ has been taken for this family of perturbation $\phi(\alpha)$, where $\Xi$ involves a region that begins from the so-called  bifurcation surface $B$ at one end and continues up to the BS boundary  portion $H$ so that it turns  spacelike when it reaches $\Xi_{1}$ at the other end. It  approaches   asymptotical flatness at  infinity. The spacetime geometry representing Buchdahl star is assumed to be linearly stable under the one parameter family of field $\phi(\alpha)$ perturbation. Note that we further work with Gaussian null coordinates at $H$, where one can write
\begin{eqnarray}\label{Eq:bs}
\int_B\delta \mathbf{Q}_\zeta(\alpha)=\frac{k}{8\pi}A_{B}(\alpha)\, ,
\end{eqnarray}
where $A_{B}$ refers to the bifurcate surface area~\cite{Sorce-Wald17}. This perturbation however does  vanish on the bifurcation surface $B$ as a consequence of hypersurface's property. Regardless of this fact, it turns out that the spacetime geometry tends to new perturbed state with $M(\alpha)$ and $Q(\alpha)$ at very late time due to the dynamical field perturbed by in-falling matter sources.

Following the perturbation Eq.~(\ref{Eq:bs}) and recalling Eq.~(\ref{Eq:first-order1}) we further derive inequality for the linear order perturbation
\begin{eqnarray}\label{Eq:lin1}
\delta M&=&\int_B[\delta \mathbf{Q}_\zeta-\zeta\cdot\bm{\Theta}(\phi,\delta\phi)]-\int_\Xi\delta \mathbf{C}_\zeta\, .
\end{eqnarray}
It turns out that the first term in the above equation vanishes at the bifurcation surface, and so it remains
\begin{eqnarray} \label{Eq:zeta1}
\int_\Xi\delta \mathbf{C}_\zeta=
\int_{H}\epsilon_{aijk }\zeta_{(t)}^{b}\left(\delta T_{b}^{a} + \textbf{A}_{b} \delta j^{a}\right)\, .
\end{eqnarray}
By imposing the condition $\Phi_{+}=-\zeta^{b}\textbf{A}_{b}\vert_{H}$  with $\int_{H}\delta(\epsilon_{aijk}j^{a})=\delta Q$, Eq.~(\ref{Eq:lin1}) takes the form as
\begin{eqnarray}
\delta M-\Phi_{+}\delta
Q = - \int_{H}\epsilon_{aijk} \zeta_{b} \delta T^{ab}\, ,
\end{eqnarray}
where the volume element is defined by $\epsilon_{aijk}=-4k_{[a} \tilde{\epsilon}_{ijk]}$ at the BS boundary  portion $H$. For that it is straightforward to obtain the null energy condition satisfying
\begin{eqnarray}
\delta T_{ab}k^{a}k^{b}\geq 0\, .
\end{eqnarray}
This, in turn, leads to the following form for linear order perturbation
\begin{eqnarray}
\label{Eq:first-order2}
\delta M-\Phi_{+}\delta Q\geq 0\, .
\end{eqnarray}

In a similar way, for non-linear  perturbation,  one obtains
\begin{eqnarray}
\delta^2 M&=&\int_B[\delta^2 \mathbf{Q}_\zeta-\zeta\cdot\delta\bm{\Theta}(\phi,\delta\phi)]-\int_\Xi\zeta\cdot\delta\mathbf{E}\delta\phi\nonumber\\&-&\int_\Xi\delta^2 \mathbf{C}_\zeta+\mathcal{E}_\Xi(\phi,\delta\phi)\nonumber\\
&=& \int_B[\delta^2 \mathbf{Q}_\zeta-\zeta\cdot\delta\bm{\Theta}(\phi,\delta\phi)]+\mathcal{E}_H(\phi,\delta\phi)\nonumber\\&-&\int_H\zeta\cdot\delta\mathbf{E}\delta\phi-\int_{H}\epsilon_{aijk}\zeta_{(t)}^{b}\left(\delta^2 T_{b}^{a} + \textbf{A}_{b} \delta^2 j^{a}\right)\nonumber\\ &=& \int_B[\delta^2 \mathbf{Q}_\zeta-\zeta\cdot\delta\bm{\Theta}(\phi,\delta\phi)]+\mathcal{E}_H(\phi,\delta\phi)\nonumber\\&+& \int_{H}\tilde{\epsilon}_{ijk} k_{a}\zeta_{b} \delta^2 T^{ab}+\Phi_{+}\delta^2
Q\, .
\end{eqnarray}
In the above equation, we used the gauge condition $\zeta^{a}\delta \textbf{A}_{a}=0$ on the  portion $H$ with the timelike Killing vector field $\zeta^{a}$ always tangent to $H$. We then apply $\delta^2 T_{\alpha\beta}k^{\alpha}k^{\beta}\geq0$ to write
\begin{eqnarray}\label{Eq:second-order2}
\delta^2 M-\Phi_{+}\delta^2
Q&=& \int_B[\delta^2 \mathbf{Q}_\zeta-\zeta\cdot\delta\bm{\Theta}(\phi,\delta\phi)]\nonumber\\&+&\mathcal{E}_H(\phi,\delta\phi)\, .
\end{eqnarray}
One can then further consider $\phi(\alpha)^{BS}$ as a perturbed field as a consequence of matter falling into the Buchdahl star. After matter is absorbed,  Buchdahl star parameters become
\begin{eqnarray}\label{MQ}
M(\alpha)= M+\alpha\delta M\, ~~\mbox{and}~~
Q(\alpha)=Q+\alpha\delta Q\, .
\end{eqnarray}
From the above equation, we shall choose $\delta M$ and $\delta Q$ as the linear order perturbations, as shown in Eq.~(\ref{Eq:first-order2}). Let us then compute rest of the terms on the right hand side of Eq.~(\ref{Eq:second-order2}) for the perturbation field $\phi^{BS}$, according to which
$\delta^2 M=\delta^2 Q_{B}=\delta E=\mathcal{E}_{H}(\phi,\delta\phi^{BS})=0$ always. Therefore, for this perturbation field we apparently have
\begin{eqnarray}
\delta^2 M-\Phi_{+}\delta^2
Q&=&\int_B[\delta^2 \mathbf{Q}_\zeta-\zeta\cdot\delta\bm{\Theta}(\phi,\delta\phi^{BS})]\, .
\end{eqnarray}
From the above equation, the term $\int_B\zeta\cdot\delta\bm{\Theta}(\phi,\delta\phi^{BS})$  vanishes since $\zeta^{a}=0$ continues to hold at the bifurcation surface $B$. Hence, for the  perturbation field, {taking into account Eq.~(\ref{Eq:bs})} the non-linear  variational inequality is given by
\begin{eqnarray}\label{Eq:non-linear}
\delta^2 M-\Phi_{+}\delta^2 Q\geq -\frac{k}{8\pi}\delta^2 A^{BS}\, .
\end{eqnarray}

Following this new version of the gedanken experiment as alluded above, we investigate  overcharging of
a nearly extremal Buchdahl star, being an analogue of extremal black hole. We recall Eq.~(5) indicating  extremality by  {$\beta^2=9/8$} and over-extremality by {$\beta^2>9/8$}. The latter means there exists no boundary radius for BS. For BH, it then exposes the singularity -- bares it naked while for BS, it would mean that the star would probably disperse  away. This is perhaps because since it is already overcharged,  repulsive contribution due to charge would override attraction due to mass.

We first explore an extremal BS; i.e.,
\begin{eqnarray}
M^2-\frac{8}{9}Q^2=0\, .
\end{eqnarray}
Note that when a particle falls into BS, its parameters would change to  $M + \delta M$ and $Q+\delta Q$. An extermal BS
can be overcharged if and only if the following inequality
\begin{eqnarray}\label{Eq:extremal}
\delta M & < &\frac{8}{9}\frac{Q }{M}\delta Q\, ,
\end{eqnarray}
holds good.
However, for an extremal BS the corresponding electromagnetic potential is
\begin{eqnarray}
\Phi_{r_{BS}}=\frac{8}{9}\frac{Q }{M}\, .
\end{eqnarray}
Thus, as mentioned earlier, when the null energy condition
is satisfied,  the linear order perturbation inequality (i.e. Eq.~(\ref{Eq:first-order2})) becomes
\begin{eqnarray}
\delta M-\frac{8}{9}\frac{Q }{M}\delta Q\geq 0\, .
\end{eqnarray}

The above two inequalities are in clear contradiction, hence  an extremal Buchdahl star can not be overcharged via the gedanken experiment.

Let us then consider a near extremal Buchdahl star and apply  the gedanken experiment afresh allowing for second order perturbations. We consider one parameter family of perturbations, $g(\alpha)$,
and write
\begin{eqnarray}\label{Eq:g1}
g(\alpha)&=&M(\alpha)^2-\frac{8}{9}Q(\alpha)^2\, .
\end{eqnarray}

Now the question is what happens when  higher order perturbations are included, would the linear order result hold good or not? For a near extremal BS, we write from Eq.~(\ref{Eq:g1}) $g(0)=M^2\epsilon^2=M^2(1-(8/9)Q^2/M^2)^{1/2}$ with $\epsilon\neq 0$ and $\alpha \neq 0$,  the function $g(\alpha)$ can be expanded up to second order in both $\epsilon$ and $\alpha$ as
 \begin{eqnarray}\label{Eq:g2}
g(\alpha)&=&M^2\epsilon^2+\left[2 M\delta M -\frac{16}{9} Q \delta Q\right]\alpha\nonumber\\&+&
\left[M\delta^2 M -\frac{8}{9}Q \delta^2 Q+\delta M^2 -\frac{8}{9}\delta Q^2\right]\alpha^2\nonumber\\
&+&O(\alpha^3, \alpha^2\epsilon,\alpha\epsilon^2,\epsilon^3)\, .
\end{eqnarray}
Here the second and third terms respectively correspond to the linear and non-linear order perturbations.

Following  \cite{Hubeny99,Sorce-Wald17}, we define the minimum possible value of $\delta M$ given by the  requirement that particle falls into the star which is given by
\begin{eqnarray}\label{Eq:optimal}
\delta M & \geq &\frac{Q}{r_{r_{BS}}}\delta Q =\frac{8}{9}\frac{Q }{M}\delta Q \left(1-\epsilon\right) + O(\epsilon^2)\, ,
\end{eqnarray}
This refers to the minimum energy (i.e. the optimal choice of linear order perturbation) for a charged particle to enter the BS.  For the above minimal energy it is straightforward to obtain linear order perturbation of $g(\alpha)$ in the following form
\begin{eqnarray}\label{Eq:f}
g(\alpha)&=& M^2\epsilon^2+2 M({8}/{9})^{1/2}\delta Q\left[\sqrt{\frac{1-\epsilon}{1+\epsilon}}-\sqrt{1-\epsilon^2}\right]\alpha \nonumber\\&=& M^2\epsilon^2-2\, ({8}/{9})^{1/2}M\delta Q\,\epsilon \,\alpha+\mathcal O(\alpha^2)\, .
\end{eqnarray}
From Eq.~(\ref{Eq:f}) it is obvious that $g(\alpha)<0$ is attainable for the linear order perturbation  if and only if the following condition is satisfied
\begin{eqnarray}\label{Eq:con_linear}
\delta Q>\frac{9}{16}\frac{M^2}{Q}\frac{\epsilon}{\alpha}=\frac{M}{2\,(8/9)^{1/2}}\frac{\epsilon}{\alpha}\, .
\end{eqnarray}
Hence, it is possible to overcharge BS by  linear order perturbation as is the case for charged black hole.

Next let us include the second order  perturbations  using the optimal choice for the linear order  Eq.~(\ref{Eq:optimal}). For the non-linear part, i.e.,  the third term of Eq.~(\ref{Eq:g2}) can be written as follows:
\begin{eqnarray}\label{Eq:f22}
&&\left[M\delta^2 M -\frac{8}{9}Q \delta^2 Q+\delta M^2 -\frac{8}{9}\delta Q^2\right]\alpha^2\nonumber\\&&=M\left[\delta^2 M -\Phi_{+}\delta^2 Q
+\frac{1}{M}\left(\left(\frac{8}{9 }\right)^{1/2}\sqrt{\frac{1-\epsilon}{1+\epsilon}}\delta Q\right)^2\right. \nonumber\\&&-\left.\frac{8}{9}\frac{\delta Q^2}{M}\right]\alpha^2=
M\left[\delta^2 M -\Phi_{+}\delta^2 Q
+\mathcal O(\epsilon) \right]\alpha^2\, .
\end{eqnarray}
Taking into account Eq.~(\ref{Eq:non-linear}) $\delta^2 M -\Phi_{+}\delta^2 Q$ in the above equation can be written as follows: %
\begin{eqnarray}\label{Eq:sec_in}
\delta^2 M -\Phi_{+}\delta^2 Q\geq -\frac{k}{8\pi}\delta^2 A^{BS}\, ,
\end{eqnarray}

with the BS boundary area $A$ and the surface gravity
$k$, as shown in Eq.~(\ref{Eq:k}). It is then straightforward to obtain the surface gravity, $k$, in the following form  \begin{eqnarray}
k &=&\frac{F'(r)}{2}|_{r=r_{+}}\,\nonumber\\&=& \frac{ M (8/9)^2}{(M+M\epsilon)^2}-\frac{ (1-\epsilon^2)M^2(8/9)^2}{(M+M\epsilon)^3}\nonumber\\&=& \frac{64}{81M}\epsilon+\mathcal O(\epsilon^2)\, .
\end{eqnarray}

Following the above procedure we then obtain the analytic form for $\delta^2 A^{BS}$ as follows:
\begin{eqnarray}
\delta^2 A^{BS}&= & \frac{9\pi}{4 \left(9\,M^2-8\, Q^2\right)^{3/2}}\Bigg[27 M^3 \left(9 \,\delta M^2-4\, \delta Q^2\right)\nonumber\\&+&9 M^2 \left(9 \delta M^2-4\delta Q^2\right) \sqrt{9 M^2-8 Q^2}\nonumber\\&+& 8 Q^2 \Big(-9  \sqrt{9 M^2-8 Q^2}\,\delta M^2\nonumber\\&+&24\,Q \,\delta M \,\delta Q+4  \sqrt{9 M^2-8 Q^2}\delta Q^2 \Big)\nonumber\\&-&324 M Q^2\delta M^2 \Bigg]\nonumber\\&=&-\frac{9\pi(1+\epsilon)}{\epsilon^3}\Bigg[2 (\epsilon -1) \sqrt{\frac{2}{\epsilon +1}-1} \sqrt{1-\epsilon^2 }\nonumber\\&+&2-  \left(4-3 \epsilon\right)\,\epsilon \Bigg]\delta Q^2 =-\frac{9\pi\left(1+\epsilon\right)}{\epsilon}\delta Q^2+\mathcal O(\epsilon)\, .\nonumber\\
\end{eqnarray}
Given the above equations for $k$ and $\delta^2 A^{BS}$, the second order inequality yields
\begin{eqnarray}\label{Eq:sec_in1}
\delta^2 M -\Phi_{+}\delta^2 Q&\geq & -\frac{k}{8\pi}\delta^2 A^{BS}\nonumber\\&=&\frac{8\left(1+\epsilon\right)}{9M}\delta \,Q^2+\mathcal O(\epsilon)\, .
\end{eqnarray}

Taking into account all the above we rewrite Eq.~(\ref{Eq:g2}) in the following form
\begin{eqnarray}\label{Eq:final}
g(\alpha)&=& M^2\epsilon^2-2\, ({8}/{9})^{1/2}M\delta Q\,\epsilon \,\alpha+({8}/{9})\delta Q^2\alpha^2\nonumber\\&+&O(\alpha^3, \alpha^2\epsilon,\alpha\epsilon^2,\epsilon^3)\nonumber\\&=&\left(M\epsilon-({8}/{9})^{1/2}\,\delta Q\,\alpha\right)^2+O(\alpha^3, \alpha^2\epsilon,\alpha\epsilon^2,\epsilon^3)\, .\nonumber\\
\end{eqnarray}
This clearly shows that $g\geq0$ always. Thus when second order perturbations are included, BS can never be overcharged and WCCC would be always respected. As for BH,  WCCC may be violated at the linear order, which would though be always overturned at the non-linear order.  Non-linear perturbations thus always favour WCCC for both BH and BS.


\section{Discussion}
\label{Sec:Discussion}

In this paper, the main aim is to extend WCCC to a non black hole compact object like BS. This question is motivated by the fact that the latter does share almost all the properties with the former~\cite{Dadhich22,Sum-Dad22}. It is indeed the most compact non black hole object \cite{Pani22}. What was then expected was that, as for BH, WCCC may be violated at the linear order which would however be restored when non-linear perturbations were included. This is precisely what we have been able to bear out for the Buchdahl star by a detailed analysis paralleling the BH case.

The one critical difference between BH and BS is that  boundary of the former is a null surface from which nothing could come out while for the latter it is timelike allowing both way crossing. What it would mean for accretion process being considered for overcharging of BS is that matter might also have some outflow. This would further hinder the overcharging effort. It therefore appears that timelike boundary of BS may in fact further work counter to the accretion process of overcharging. This would have been very pertinent, had the case been otherwise; i.e., accretion process  helped overcharging. 

One may ask that this analysis could have been carried out at  any radius $r$. No, that couldn't have been the case because the question of overcharging arises only when the extremal limit is defined. The extremal limit is defined only for BH and BS as characterized respectively by $\Phi(r) = 1/2, 4/9$. It is however true that the analysis proceeds as if the boundary is null. The pertinent point is that its non-null character does not go contrary to the result, instead it goes in line. It should however be appreciated that even though BS boundary is not null, it is very close to the BH horizon and hence for all practical purposes it is almost as compact an object as BH. It should therefore be not surprising if it shares the properties with BH.

There is one very distinct difference between the linear and non-linear case. For the former, the WCCC violation requires the condition Eq.~(\ref{Eq:con_linear}) while there is no such condition as shown in Eq.~(\ref{Eq:final}) for the non-linear case. That is, WCCC would always be respected without any constraints on the parameters.

Another related question is, could a non-extremal BS be extremalized; i.e., converting non-extremal into extremal.  This is not possible for BH as shown in ~\cite{Dadhich97}. This was because as extremality is approached, the parameter window for infalling particles pinches off prohibiting extremalization. It would therefore be expected that the same should be the case for BS which we would next intend to examine in a separate investigation.

Like many other properties, BS shares yet another property with BH, in the context of the weak cosmic censorship conjecture, which may be violated at the linear order perturbations but is always restored at the non-linear order.

\section*{Acknowledgments}
This work is supported by the National Natural Science Foundation of China under Grants No. 11675143 and No. 11975203, the National Key Research and Development Program of China under Grant No. 2020YFC2201503. N.D. wishes to acknowledge the support of CAS President's International Fellowship Initiative Grant No. 2020VMA0014.

\appendix
\bibliographystyle{JHEP}
\bibliography{gravreferences,reference}

\end{document}